\documentclass[ aps, amsmath,amssymb, pra]{revtex4-2}

\usepackage{graphicx}
\usepackage{dcolumn}
\usepackage{bm,color}

\begin{document}

\preprint{APS/123-QED}

\title{Quantum transport signature of strain-induced scalar and pseudo-vector potentials in a crenellated hBN-graphene heterostructure}

\author{Romaine Kerjouan}
\altaffiliation{1Laboratoire de Physique de l’Ecole Normale Supérieure, ENS, Université PSL, CNRS, Sorbonne Université, Université de Paris, Paris, France.}
\author{Michael Rosticher}
\altaffiliation{1Laboratoire de Physique de l’Ecole Normale Supérieure, ENS, Université PSL, CNRS, Sorbonne Université, Université de Paris, Paris, France.}
\author{Aurélie Pierret}
\altaffiliation{1Laboratoire de Physique de l’Ecole Normale Supérieure, ENS, Université PSL, CNRS, Sorbonne Université, Université de Paris, Paris, France.}
\author{Kenji Watanabe}
\altaffiliation{2Research Center for Functional Materials, National Institute for Materials Science, 1-1 Namiki, Tsukuba 305-0044, Japan}
\author{Takashi Taniguchi}
\altaffiliation{3International Center for Materials Nanoarchitectonics, National Institute for Materials Science, 1-1 Namiki, Tsukuba 305-0044, Japan}
\author{Sukhdeep Dhillon}
\altaffiliation{1Laboratoire de Physique de l’Ecole Normale Supérieure, ENS, Université PSL, CNRS, Sorbonne Université, Université de Paris, Paris, France.}
\author{Robson Ferreira}
\altaffiliation{1Laboratoire de Physique de l’Ecole Normale Supérieure, ENS, Université PSL, CNRS, Sorbonne Université, Université de Paris, Paris, France.}
\author{Daniel Dolfi}
\altaffiliation{4Thales Research \& Technology, Thales Group, Palaiseau, France}
\author{Mark Goerbig}
\altaffiliation{5Laboratoire de Physique des Solides, CNRS, Université Paris-Saclay, Orsay, France}
\author{Bernard Plaçais}
\altaffiliation{1Laboratoire de Physique de l’Ecole Normale Supérieure, ENS, Université PSL, CNRS, Sorbonne Université, Université de Paris, Paris, France.}
\author{Juliette Mangeney}
\altaffiliation{1Laboratoire de Physique de l’Ecole Normale Supérieure, ENS, Université PSL, CNRS, Sorbonne Université, Université de Paris, Paris, France.}
\email{juliette.mangeney@phys.ens.fr}

\date{\today}
             
\begin{abstract}
The sharp Dirac cone of the electronic dispersion confers to graphene a remarkable sensitivity to strain. It is usually encoded in scalar and pseudo-vector potentials, induced by the modification of hopping parameters, which have given rise to new phenomena at the nanoscale such as giant pseudomagnetic fields and valley polarization. Here, we unveil the effect of these potentials on the quantum transport across a succession of strain-induced barriers. We use high-mobility, hBN-encapsulated graphene, transferred over a large (10x10 $\mu$m$^{2}$) crenellated hBN substrate. We show the emergence of a broad resistance ancillary peak at positive energy that arises from Klein tunneling barriers induced by the tensile strain at the trench edges. Our theoretical study, in quantitative agreement with experiment, highlights the balanced contributions of strain-induced scalar and pseudo-vector potentials on ballistic transport. Our results establish crenellated van der Waals heterostructures as a promising platform for strain engineering in view of applications and basic physics.
\end{abstract}

\maketitle


\section{\label{sec:level1}Introduction}

Graphene exhibits high mechanical flexibility and remarkable electronic properties, making it an ideal platform for strain engineering to explore novel fundamental phenomena and to realize straintronic devices \cite{Zhai2019}. Elastic strain gives rise to two primary effects in the low-energy band structure of graphene: shifts in the energy of the Dirac point, which is typically incorporated as a scalar potential and shifts in the momenta of the Dirac cones in opposite directions for the two valleys, often described by a pseudo-vector potential \cite{Kitt2012,mcrae2019graphene,Botello-Mendez2018,MPFG1, MPFG2, pereira2009tight, pereira2009strain, Guinea2008}. Strained graphene has been the subject of a large number of studies, revealing fascinating physical phenomena that depend mainly on whether the strain is non-uniform or not. Non-uniform strains over typically a few nanometers create in graphene giant pseudo-magnetic fields that can reach several hundreds of Tesla \cite{levy2010strain}, giving rise to the pseudomagnetic quantum Hall effect \cite{Liu2018}, and the proposals of
valley splitting topological channels for chiral fermions \cite{Hsu2020} or the appearance of superconductive state \cite{Uchoa2013,Xu2018}. Such non-uniform strain has been obtained on localized curved structures such as wrinkles or bubbles \cite{Reserbat-Plantey2014,Liu2018, Goldsche2018} and, recently, on a large scale using twisted multilayers \cite{Liu2018} and substrate nanopatterning \cite{Jiang_2017, Hsu2020}, making it possible to study the pseudo-magnetic field using magneto-transport measurements \cite{Liu2018,Ho2021}. Uniform strain, on the other hand, creates large scalar potentials that modify the graphene work function \cite{he2015tuning}. By achieving uniform strain in mesoscopic graphene devices, strain-induced scalar potentials have been probed using transport experiments \cite{wang2020global,Hiroki2015}. In spite of intensive work, the investigation of the quantum transport of Dirac fermions through a network of uniform strained barriers remains elusive in graphene. \\

Probing the quantum transport of relativistic electrons in graphene through a strained barrier is of considerable interest as it differs remarkably from transport across an electrostatic barrier, such as a p-n-p junction. For instance, Dirac fermions undergo Klein tunneling across an electrostatic barrier, and those at normal incidence, constrained to retain their transverse momentum, $k_{\bot}=0$, and forbidden to scatter directly backwards, penetrate the barrier with unit probability. In contrast, the pseudo-vector potential in strained graphene, which is absent in electrostatically defined n-p-n junctions, shifts the transverse momentum, $k_{\bot}\rightarrow k_{\bot}-A_{\bot}$, of electrons inside the barriers, affecting the quantum transport of electrons which is sensitive to both energy and momentum conservation. For instance, partial reflections for carriers normally incident on the strained junction are predicted. This effect, which is specific to ballistic graphene, is irrelevant in diffusive graphene where momentum recoil is supplied by impurities. \\

 Furthermore, it is worth pointing out that quantum transport of electrons in graphene through a network of uniform strained barriers is a somewhat unusual situation in which the pseudo-vector potential has a physical significance. This would not be the case in systems with a homogeneous strain, in which the global shift $k_{\bot}\rightarrow k_{\bot}-A_{\bot}$ of the positions of the Dirac points inside the first Brillouin zone has no physical consequence and could be compensated by a global gauge choice of the vector potential. Such a global gauge choice is prohibited for a succession of strained and unstrained graphene regions. On the other hand, pseudo-magnetic fields associated with the curl of the vector potential are expected to play no essential role in the quantum transport of electrons through a network of uniform uniaxial strained barriers. Indeed, such pseudo-magnetic fields and the associated Landau quantization require a broken inversion symmetry between the two graphene sublattices, as it is the case in the particular strain patterns with a $2\pi/3$-rotation symmetry observed in scanning-tunneling-spectroscopic experiments \cite{levy2010strain}. Apart perhaps from few localized impurities, the pseudo-magnetic field therefore has a negligible contribution to quantum transport across a uniaxial strained graphene barrier, while the pseudo vector-potential contribution, which vanishes in diffusive graphene, is expected to dominate the ballistic transport. Quantum transport across a uniform strained barrier can thus serve as a highly sensitive probe of strain-induced pseudo-vector potentials, which have been little studied compared to pseudo-magnetic fields. \\

Here, we report on a mechanically-robust strained graphene mesoscopic device made of a crenellated Van der Waals hBN-graphene heterostructure. Our approach relies on transferring a high-quality graphene layer protected by a thin hBN layer, on a crenellated hBN substrate to create a periodic network of strained and unstrained regions. The strained regions act as barriers for ballistic electrons and thus, in the device, electrons propagate through a succession of strain barriers. We probe the Klein tunneling properties of ballistic electrons across the strain-induced barriers using low-temperature transport measurements, and model them by combining elasticity and tight-binding theories with Dirac fermion optics and Landauer-Buttiker scattering approaches. We unveil a signature of tensile strain in the form of a large and broad ancillary resistance peak at a positive energy. We show that peak amplitude and shape result from balanced contributions of scalar and pseudo-vector potentials in the Klein tunneling strained barriers. 

\section{\label{sec:transport}Mesoscopic transport across a crenellated hBN-graphene heterostructure}

The device (Fig.1a) is made of a (LxW=12x8 $\mu$m) exfoliated monolayer graphene protected by a thin hBN layer and transferred onto a 62 nm-thick hBN layer nanopatterned in periodic trenches of 35 nm depth, resting on a silicon substrate with 500 nm oxide (SiO$_{2}$) used as a back gate dielectric (see Fig.1b and Methods for fabrication process). Due to strong Van der Waals interactions, the hBN/graphene heterostructure adheres to the crenellated hBN substrate at the top and bottom of the slot, as shown in the AFM image (Fig. 1c). This anchors minimally strained regions concentrating the tensile strain in the suspended graphene region connecting the two anchored regions; its length is determined by the balance between the out-of-plane component of the tension and the hBN adhesion force. Thus, our approach yields 300 nm long regions of  unstrained graphene on top and 400 nm long regions in the bottom, and typically 150 nm-long regions of tensile strained graphene along the trenches. From basic geometrical considerations, we can estimate the strain in our device to be a few percent. We infer high electronic mobility in the graphene on the top of the crenellation as the hBN/graphene heterostructure is anchored to pristine hBN\cite{dean2010boron} and in the tensile strained region because the graphene is suspended \cite{Bolotin2008}. In contrast, in the bottom of the crenellation, the quality of the graphene is degraded by the roughness of the hBN following the etching step, which suppresses high carrier mobility. Quantum transport across a N-crenellated device can thus be regarded as the sum N-diffusive sections and 2N strained-junctions in series. Raman spectroscopy in Fig.1d confirms the relatively high quality of the graphene as the average intensity of the 2D peak, I$_{2D}$, is 2.5 times higher than that of the G peak, I$_{G}$, and the average intensity of the D peak, I$_{D}$, remains 3 times lower than that of the G peak \cite{ferrari2013raman}. Note that the I$_{2D}$/I$_{G}$ ratio, which can be considered as a measure of graphene quality, follows the same periodicity as the crenellation with high ratio bands at the top regions and low ratio bands at the bottom regions. This behavior supports our assumption of a high quality graphene on the top of the crenellation where the hBN has only been exfoliated and a low quality graphene on the bottom of the crenellation where the hBN has been etched prior to graphene transfer. In addition, positions of peaks 2D and G over the crenellated device are clearly linearly correlated (see Fig.1d), the slope of the linear regression is 2.46 which is typical of presence of strain in graphene \cite{lee2012optical, wang2015uniaxial} confirming the presence of strained regions in crenellated hBN-encapsulated graphene. The whole device includes N=10 periodic crenellations (period l=1 $\mu$m length, see Fig.1e) contributing additively to the total resistance. They are separated from the source and drain “edge contacts” by 2 unstrained graphene regions of 1 µm length, which add a diffusive contribution to the total resistance \cite{wang2013one}. The large thickness of the SiO$_{2}$ oxyde with regards to the hBN trench depth results in a modest variation of the gate capacitance within the whole structure, between C$_{bottom}$=6.3$\times$10$^{-9}$ F.cm$^{-2}$ and C$_{top}$=5.9$\times$10$^{-9}$ F.cm$^{-2}$, i.e. $<$7 percent. Therefore, the modulation of the carrier density by the global back gate is relatively uniform over the entire device.\\

\begin{figure}[ht]%
\centering
\includegraphics[width=1\textwidth]{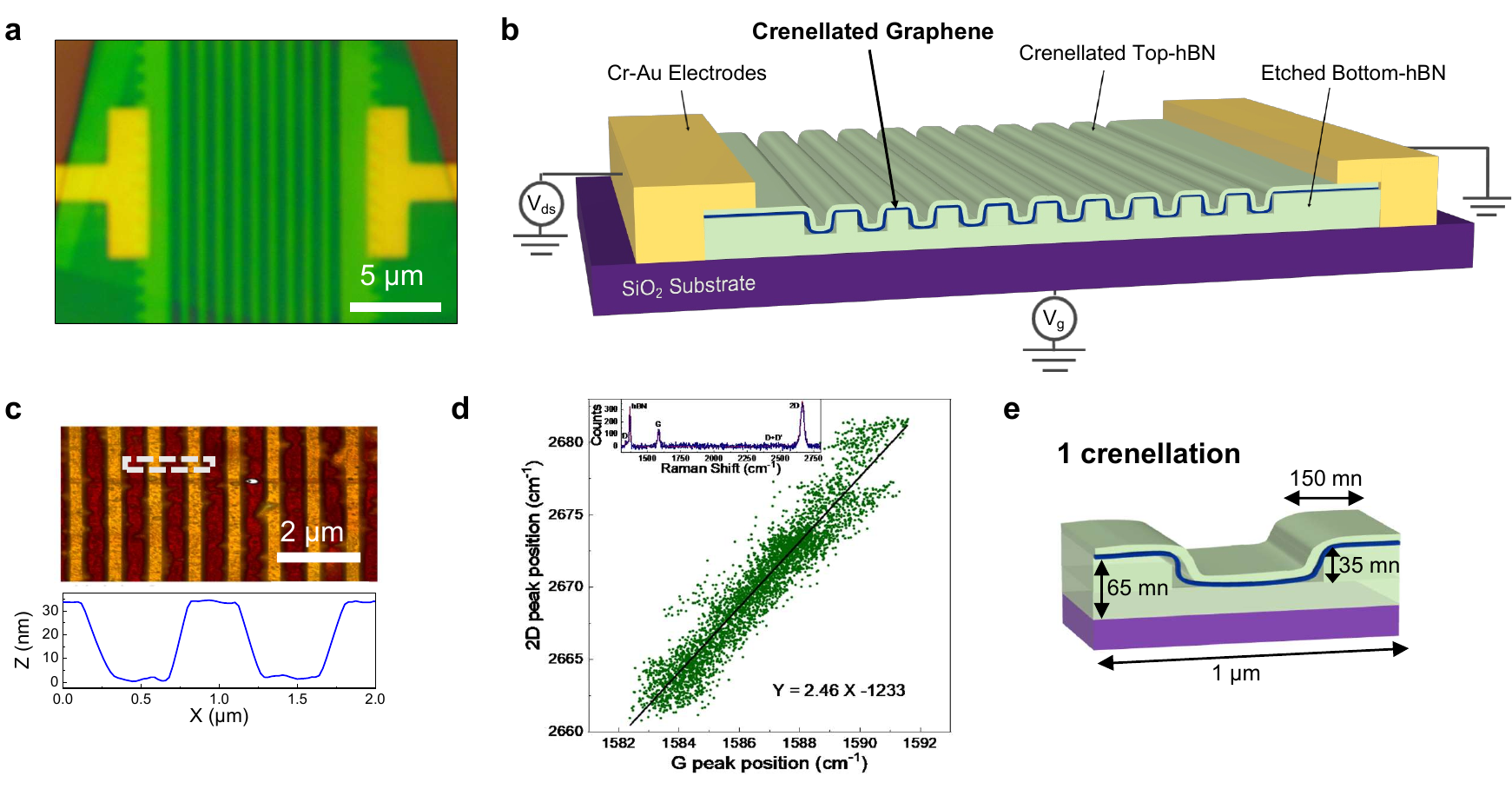}
\caption{\textbf{Crenellated hBN/graphene heterostructure}:  Optical image (a) and schematic drawing (b) of the crenelated hBN/encapsulated graphene transistor. c) Atomic force microscopy image showing that the hBN/graphene heterostructure sticks to the crenellated hBN substrate at the top and bottom of the slot (upper). Line-cut profile displaying that the amplitude of the trenches depths is typically 35 nm (lower). d) In green, positions of 2D peaks as a function of positions of G peaks overall the corrugated area demonstrating clear correlation. Linear regression, in black, indicates a slope of 2.46 typical of the presence of strain in graphene. Inset : typical Raman spectrum of our device, blue line, experimental data, red dashed line, fit with 5 Lorentzians that are characteristic of graphene - peaks D, G, D + D', 2D - and of hBN - peak hBN. e) Schematic drawing of 1 crenelation that includes a two 150 nm long regions of high-quality unstrained graphene on top and one 400nm long region in the bottom and two $\sim$ 150nm-long regions of tensile strained graphene along the trenches, 
}\label{fig1}
\end{figure}

The signature of strain on transport is displayed in Fig.2a, which reports the low-bias differential resistance R$_{ds}$ of the device as a function of gate voltage V$_{g}$ for different temperatures ranging from 4.2 K to 100 K. The R$_{ds}$-V$_{g}$ transport-characteristics reveal a broad and large ancillary resistance peak at positive energy (chemical potential $\mu_{peak}$=+163 meV for V$_{g}$= 62 V) in addition to the usual sharp charge neutrality peak (at V$_{g}$=2 V). The Dirac peak (illustrated by the dashed blue curve in Fig.2a) corresponds to the charge neutrality point of the unstrained graphene region and its position close to 0 V confirms the high quality of the graphene layer. The broad ancillary peak (illustrated by dashed red curve in Fig.2a), which does not exist in flat graphene, has an amplitude comparable to that of the Dirac peak and a high energy position, suggesting the presence of large potential barriers (~$\mu_{peak}$) induced by the tensile strains in the device. We notice that the ancillary peak shape is in itself a signature of strain as it significantly differs from standard Dirac peaks \cite{huang2020ultra} or  Dirac Fermion reflector plateaus\cite{Wilmart_2014, Graef2019}. Moreover, we observe modulations, more visible in the gate voltage range around to the ancillary peak, that are unexpected in flat graphene transistors. These modulations can be considered as hints of ballistic transport, as they are reminiscent of the electronic Fabry-Perot oscillations observed in electrostatic barriers \cite{Young2009}. Their characteristic energy separation is  $\Delta E\approx$10 meV corresponding to an effective cavity length $L_{cavity}=2 \pi \hbar v_{F}/\Delta E$ = 360 nm \cite{Rickhaus2013}. This length is consistent with the geometric length of the unstrained high-quality graphene at the top ($\sim$300 nm) of the crenellation. We therefore attribute the observed oscillations to Fabry-Perot interferences between two strain-induced barriers (the regions of tilted graphene), with the top flat graphene regions being considered as cavities for high-mobility carriers and strain junctions as efficient barriers. We can exclude Fabry-Perot oscilations within the barrier due to the length mismatch and the robustness of oscillations at the ancillary peak.

The quantum nature of resistance oscillations is corroborated by the temperature effect, which shows strong oscillation damping as the temperature increases. Blurring of interferences resulting from quantum transport by the thermal broadening of the electrons impinging on the barriers is expected to follow an exponential suppression given by $\exp(-4\pi^2k_{B}T/\Delta E)$ with k$_{B}$ the Boltzmann constant and $T$ the temperature \cite{Deprez2021}. Considering our characteristic energy separation, a 95$\%$ drop in interference visibility corresponds to a thermal energy of k$_{B}$T=3 meV, i.e. T=35K, which is in excellent agreement with our data. This observation confirms that the observed conductance oscillations arise from quantum interferences of electrons propagating ballistically between strain barriers. \\

To get deeper insight in the transport of ballistic electrons through strained graphene barriers, we measure the evolution of the R$_{ds}$-V$_{g}$ characteristics for different bias V$_{ds}$ -from 0 mV to 120 mV- at low temperature (4.4 K). With increasing bias, we observe in Fig.2b that R$_{ds}$ decreases and the Fabry-Perot oscillations vanish and cancel above 30 mV. Comparatively, the behavior of R$_{ds}$ with temperature and bias are thus very similar. More quantitatively, we report on Fig.2c, R$_{ds}$ as a function of the bias drop per micrometer (black symbols) and of the thermal energy (red symbols) close to Dirac point (V$_{g}$=2V) and at the ancillary peak (V$_{g}$=62 V). The very good overlap between the two trends indicates that the device can be considered as 1 $\mu$m-long elements associated in series and thus supports our interpretation of the voltage drop in terms of N crenellations that contribute additively to the mesoscopic transport in our device. The differential resistance map as a function of V$_{g}$ and V$_{ds}$ is presented in Fig.2d. This 2D plot highlights the strength and robustness of the ancillary peak, attributed to propagation of ballistic electrons through the multiple strained graphene barriers. 

\begin{figure}[ht]%
\centering
\includegraphics[width=0.9\textwidth]{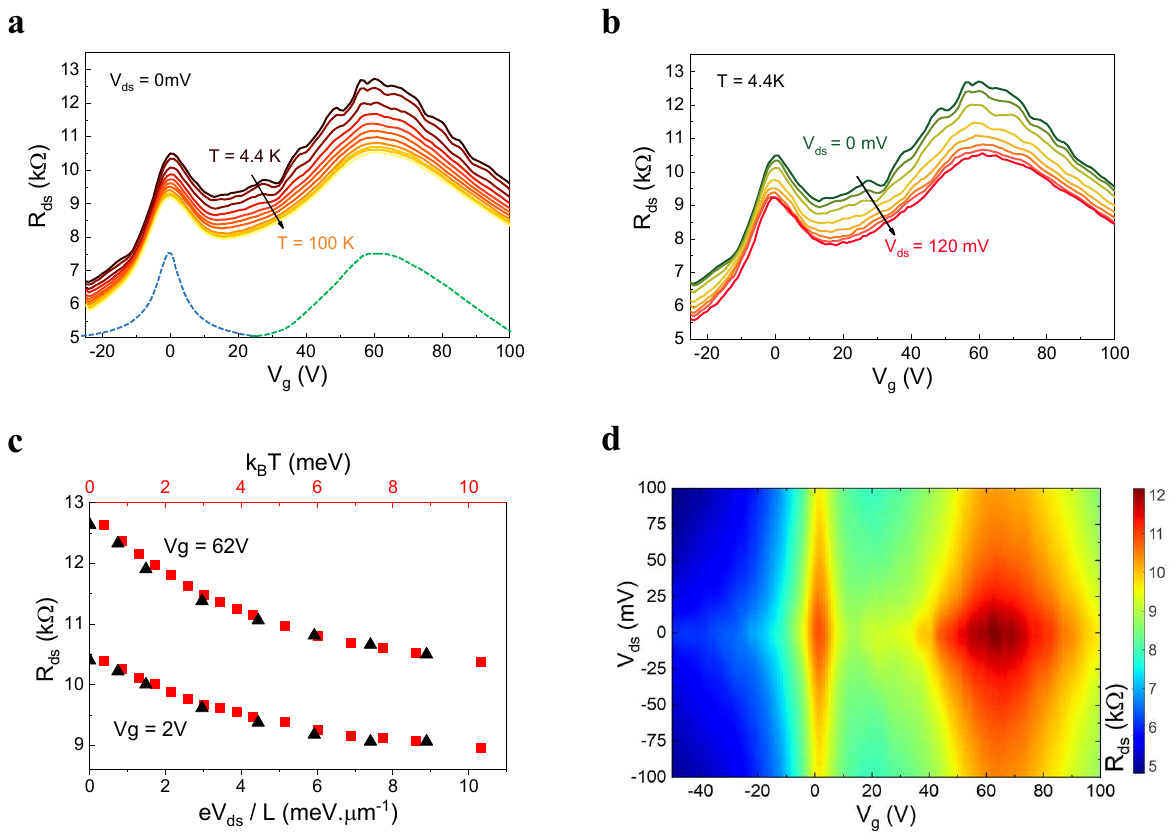}
\caption{\textbf{Low bias transport measurements} a) Low-bias resistance $R_{ds}$ measured at V$_{ds}$=10mV for different temperatures ($T =$ 4.4 K, 10 K, 20 K, 30 K, 40 K, 50 K, 60 K, 70 K, 80 K, 90 K, 100 K). We observe the emergence of a broad ancillary resistance peak at high V$_{g}$ (illustrated by the dashed green curve) in addition to the usual Dirac peak (illustrated by the dashed blue curve ). b) Evolution of R$_{ds}$-V$_{g}$ characteristics at T = 4.4 K for different bias ($V_{ds} =$ 0 mV, 10 mV, 20 mV, 40 mV, 60 mV, 80 mV, 100 mV and 120 mV). c) R$_{ds}$ as a function of the bias drop per micrometer (black symbols) and of the thermal energy (red symbols) close to Dirac point (V$_{g}$=2V) and at the ancillary peak (V$_{g}$=62 V). d) Map of differential resistance $R_{ds}$ in function of the gate voltage $V_{g}$ and of the bias voltage $V_{ds}$ at T = 4.4 K. }
\label{fig2}
\end{figure}

\section{\label{sec:theory}Theoretical description of electronic transport across a strained graphene barrier}

In the following we develop a theoretical description of quantum transport in this crenellated geometry combining elasticity and tight-binding theories, with Dirac fermion optics and Landauer-Buttiker scattering approaches. More specifically, we calculate the transmission and conductance through a single strained graphene barrier surrounded by unstrained graphene regions (more details in Supplementary). Let us start by calculating the electronic properties of strained graphene. We assume that stress is uniaxial and colinear with the direction of electronic transport. We introduce the uniaxial strain tensor in the graphene lattice coordinates, $\bar{\bar{\varepsilon}}\left(\varepsilon,\theta\right)$, with $\varepsilon$ the longitudinal deformation and $\theta$ the angle between the strain direction and the zigzag crystallographic direction of the graphene sheet. A first effect of uniaxial strain on the graphene band structure is a shift in energy, independent of the valley, which arises from changes in the next-nearest-neighbor hopping parameter \cite{manes2007symmetry,suzuura2002phonons,choi2010effects}. This effect can be described by an effective scalar potential,  $\Delta V_{strain}=g_{strain}\varepsilon(1-\sigma)$, where g$_{strain}\approx$ 3 eV \cite{guinea2010generating, sloan2013strain, schneider2015local} and $\sigma$=0.165 is the Poisson’s ratio \cite{blakslee1970elastic, pereira2009tight}). In addition, the uniaxial strain modifies the inter-atom distances between nearest neighbor, $\delta_i^\prime$, and as a consequence, the hopping amplitudes among neighboring sites change as $t=t_{0}\exp(-\beta(\lvert\overrightarrow{{\delta_i^\prime(\varepsilon,\theta)} \rvert}/a-1)$, with $\beta$=3.37  \cite{pereira2009strain}, t$_{0}$=2.7 eV and a=0.24 nm the lattice constant. Figure 3a shows the calculated low-energy dispersion and Fermi circle around Dirac point for valley K and K' in strained graphene with $\varepsilon=2\%$ and a typical $\theta$=0°. Each Dirac cone of strained graphene, K$_{i}$, shifts away from its unstrained high-symmetry position by a wave vector $\mathbf{\Delta q}_{Ki}(\varepsilon,\theta)$. Note that the 3 couples of cones go through a different displacement but only one of them belongs to the first Brillouin zone and then takes part in electronic transport to be considered here. Formally, the strain-induced momentum displacements of Dirac cones have the same effect as the application of a pseudo-vector potential, $\mathbf{A}_{K/K\prime}$, which shifts the Dirac cones away from the K and K’ points in opposite directions in the reciprocal space ($\eta =+$ for the K valley and $\eta =-$ for the K' valley). The low-energy Hamiltonian at each Dirac cone can thus be written as:
\begin{equation}
H=v_F\sigma\left(\mathbf{p}+e\mathbf{A}_{K/K\prime}\right)+\Delta V_{strain} \text{ with } \mathbf{A}_{K/K\prime}=\frac{-\eta\hbar\mathbf{\Delta q}_{K/K\prime}(\varepsilon,\theta)}{e}\
\label{hamiltonian}
\end{equation}
and v$_{F}$ the Fermi velocity in graphene, $\sigma$ the pseudo-spin operator and $\textbf{p}=\hbar \textbf{k}$ the momentum operator. Note that we disregard changes in the magnitude and isotropy of the Fermi velocity, which is a fair approximation at moderate strain \cite{pereira2009tight}. We conclude that the effect of strain on graphene leads to both a scalar potential $\Delta V_{strain}$ and a vector potential $\mathbf{A}_{K/K\prime}$, which turn out to be equally important in the analysis of our transport measurements. \\

\begin{figure}[ht]%
\centering
\includegraphics[width=0.83\textwidth]{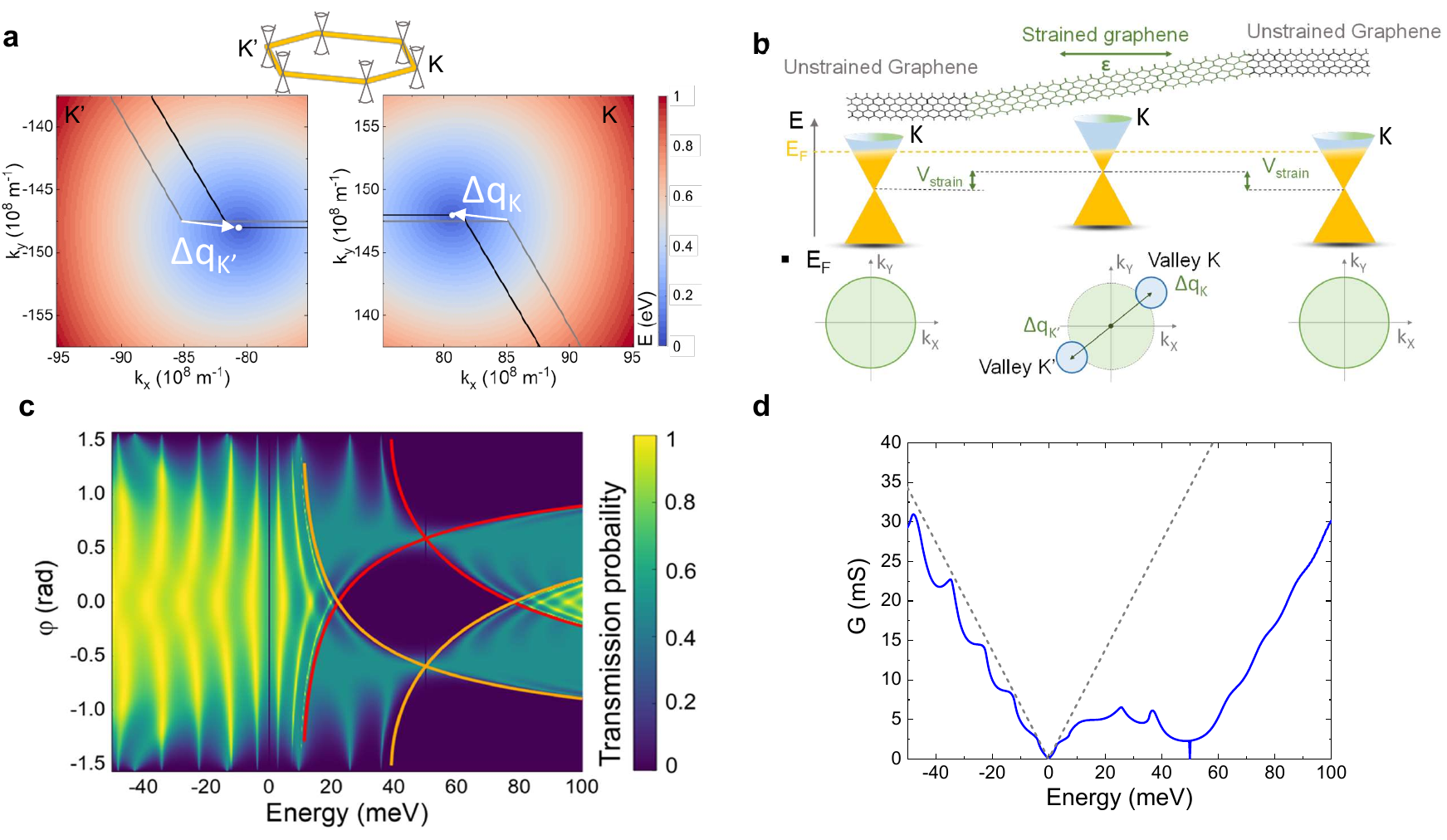}
\caption{\textbf{Transport across a barrier of strain graphene} a)  Low-energy electronic dispersions for valley K and K' for a uniaxial strain $\varepsilon = 2\%$ applied in direction $\theta = 0^\circ$ (the angle between the strain direction and the zigzag crystallographic direction of the graphene sheet). When compared to unstrained graphene, Dirac cones are shifted by $\Delta q_{K/K'} $ in opposite directions for valleys K and K'.  b) Schematic of a barrier of strained graphene - presence of an effective scalar potential $\Delta V_{strain}$ and of an effective vectorial pseudo potential $\mathbf{A}_{K/K'}$ of opposite direction according to electron valley. c) Transmission probability ($T = \vert t \vert^2$) through a 150 nm long strain barrier with uniaxial strain  $\varepsilon = 2\%$ in direction $\theta = 0 ^\circ$ as a function of $E$, the electron energy and $\varphi$, its incidence angle on the barrier. The effective pseudo potential induces changes in momentum conservation. Red lines correspond to the limits of authorized incident angles for valley K and the orange ones for valley K'. d) The ballistic conductance G for the same barrier with $8 \mu m$ width in function of the electron energy. There are two minima, the first one at $E = 0$ is due to the zero electronic density of undoped graphene and the second one at $E = \Delta V_{strain}$ is due to the decrease of authorized angles at this energy.}
\label{fig3}
\end{figure}

We now compute the transmission probability of an electron through a strained graphene barrier, as illustrated in Fig. 3b, by continuity of the electronic wave functions at the strained-unstrained interfaces \cite{pereira2009strain,fogler2008pseudomagnetic}. Figure 3c shows the calculated transmission probability T($E,\phi$) through a 150 nm long strained barrier for $\varepsilon=2\%$ and $\theta$=0° as function of the electron's energy $E$ and its incidence angle $\phi$ upon the barrier. Note that the momentum displacement parallel to the direction of transport in strained graphene, $\Delta $q$
^{\|}$, generates only an unimportant global phase on the wave functions. By contrast, the transverse component $\Delta q
^{\bot}$, encoding the vector potential A$^\bot$ and involved in the momentum conservation at the strained-unstrained interfaces, directly affects the Fresnel coefficients of electronic transport. To cross the barrier, the incident angle of the electron has to obey the modified Snell-Descartes law (see Supplementary material):
\begin{equation}
\frac{ev_{F}A_{K/K\prime}^{\bot}}{\lvert E\rvert}+\frac{\lvert E-\Delta V_{strain} \rvert} {\lvert E\rvert} \geq \sin\phi\geq\frac{ev_{F}A_{K/K\prime}^{\bot}}{\lvert E\rvert}-\frac{\lvert E-\Delta V_{strain} \rvert} {\lvert E\rvert}
\label{new_Fresnel_law}
\end{equation}
Equation \ref{new_Fresnel_law} highlights the respective roles of $\Delta V_{strain}$ and A$_{K/K\prime}^{\bot}$ in the angular limits of the total internal reflection. In particular, the angular transmission window is pinched for E=V$_{strain}$ at a finite angle entirely prescribed by $ev_{F}A_{K/K\prime}^{\bot}/\Delta V_{strain}$. This strain-induced lifting of valley degeneracy exists in principle but gives rise to non-local effects which are lost in our two-terminal experiment. In general, one can distinguish from equation \ref{new_Fresnel_law}, two limiting cases: i) the vector-potential dominated case ($\lvert ev_{F}A_{K/K\prime}^{\bot}/\Delta V_{strain}\rvert\gg 1$), where tunneling across the barrier is cancelled, and graphene becomes insulating in a broad range of energy, making it an effective on/off transistor \cite{cao2012strain}; ii) the scalar potential dominated case ($\lvert ev_{F}A_{K/K\prime}^{\bot}/\Delta V_{strain}\rvert\ll 1$), which is reminiscent of a conventional p-n junction \cite{zhai2019electron, wan2021valley}. An asset of our device is that it lies in the intermediate regime where both effects contribute constructively. It is important to point out that, in this intermediate regime, quantum transport across strained barriers differs significantly from that through an electrostatic barrier, with notably a non-zero reflection for carriers normally incident  ($\phi =0$) on strained barriers due to the pseudo-vector potential in the strained region (see Fig. 3c). Also, the resonance condition in the barrier depends not only on the barrier length and the scalar potential V$_{strain}$ similarly to the case of an electrostatic barrier, but also on the vector potential $\mathbf{A}_{K/K\prime}$, which is given by $\epsilon$ and $\theta$ and on the valley, making it very different from that of an electrostatic barrier.\\

Finally, we calculate the conductance G of electrons in graphene propagating across a strained graphene barrier at zero bias and low temperature from electron transmission probability using Landauer-Buttiker formalism (see Supplementary). Figure 3d reports the conductance of a single strained graphene barrier of 8 µm large and 150 nm long with $\varepsilon=2\%$ and a $\theta$=0°. G presents the expected minimum at the charge neutrality point ($E$=0) as well as a second minimum at $E$=$\Delta V_{strain}$ due to few incident angles allowed for crossing the strain barrier around this energy. \\

\begin{figure}[ht]%
\centering
\includegraphics[width=1\textwidth]{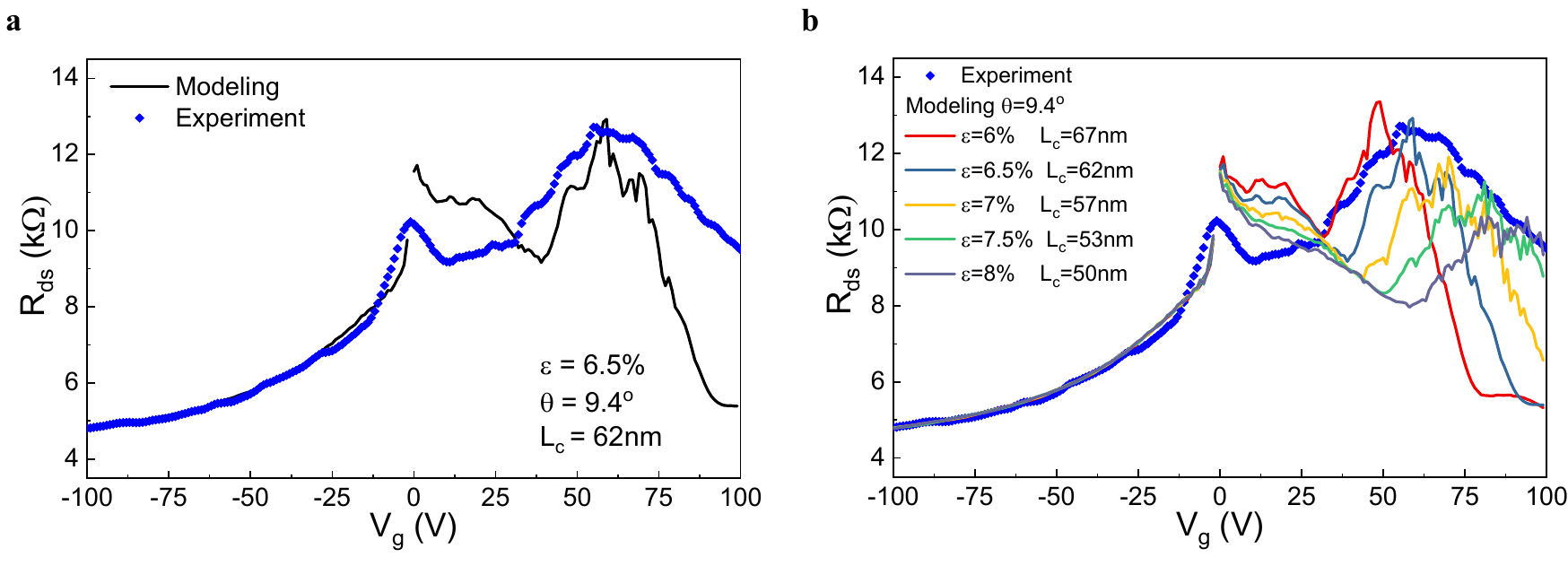}
\caption{\textbf{Quantitative comparison between experiment and theory} a) Comparison between measured $R_{ds}$ (blue symbols) and calculated $R_{ds}$ (black line) in function of $V_g$ at zero bias and $T =$4.4 K. Quantitative agreement is found for $ \varepsilon = 6,5\%$, $L_{strain} =$ 62 nm, the length of the barrier and $\theta =$ 9.4$^\circ$, the angle between the transport direction and the zigzag crystallographic direction of the graphene. b) Measured $R_{ds}$ (blue symbols) and calculated $R_{ds}$ (solid lines) in function of $V_g$ at zero bias and $T =$4.4 K for $\varepsilon$ ranging from de 6 to 8$\%$ and $L_{strain}$ ranging from 50 nm to 67 nm.}
\label{fig4}
\end{figure}

\section{\label{sec:citeref}Origin of the ancillary resistance peak}
Finally, for direct comparison with experimental data, we compute the resistance of the whole strained graphene device by calculating the resistance of a single crenellation which consists in two strained barriers and multiplying this resistance by a factor N=10 for additive contribution of the N crenellations in series. Note that we incorporate the unhomogeneous doping of the corrugation induced by its topography. Also, we add the diffusive contribution in unstrained graphene regions, calculated from fits of the charge neutrality point peak according to $R=R_{c}+L/(W\lvert n\rvert e\mu)$, with R$_{c}$ the contact resistance, $n$ the charge density of the graphene and $\mu$ the electron mobility over the entire structure, which includes the high-mobility regions on top of the crenellations and on suspended tensile graphene as evidenced by the observation of Fabry-Perot oscillations in the R$_{ds}$-V$_{g}$ characteristic and the low-mobility regions on bottom of the crenellations(see Methods) \cite{Ashcroft1976}. \\

In Fig. 4a, we compare the measured resistance with the calculated resistance, taking $\varepsilon$, $\theta$ and L$_{barrier}$ (the length of the strain barriers) as adjustable parameters. From the position of the ancillary resistance peak, we deduce $\Delta V_{strain}$=163 meV corresponding to a longitudinal deformation $\varepsilon$=6 \%, consistent with geometrical considerations. From the magnitude of the ancillary peak, we extract $\theta$=9.4° and L$_{barrier}$=62 nm. The quantitative agreement between theory and experiment demonstrates a near-equal contribution of both scalar and pseudo-vectorial potentials with ev$_{F}$A$_{\bot}$=-126 meV. We deduce at E=$\Delta V_{strain}$ an authorized transmission angle of 57° for valley K and -57° for K’ valley, which highlights a strong valley separation. However, we observe that the calculated ancillary resistance peak is narrower than experimental data. We attribute this broadening to the strain heterogeneity along a single barrier and the dispersion among the N crenellations in series, which are both incipient to our fabrication method. Nevertheless, the dispersion of strain remains modest as a strain heterogeneity $\varepsilon$=6—9 \% is sufficient to account for the broadening of the ancillary peak for $\theta$=9.4° as observed in Fig. 4b. This theoretical analysis provides a quantitative description of strain-barrier effects in the transmission of ballistic electrons and fully elucidates the origin of the ancillary resistance peak reported here. \\

We now discuss additional contributions to the observed phenomena.  As we have already mentioned in the introduction, strain in the present geometry can be described within a pseudo vector potential. However, because of the uniaxial character of the strain field, which preserves the equivalence between the two triangular sublattices of graphene, there is no curl expected to arise that could give rise to Landau-level quantization. The latter has indeed been shown to be relevant in the case of a triangular strain field that breaks the sublattice symmetry \cite{levy2010strain}. In the present device, similar localized strain fields could be present in the case of some localized disorder. However, unless this disorder breaks time-reversal symmetry, it is not expected to yield significant deviations of the electrons from their trajectories. Indeed, a possible deviation of an electron's trajectory in a particular valley should be compensated by that of an electron in the other valley, as a consequence of time-reversal symmetry. Finally, strain effects in the low-energy band structure of graphene also change the magnitude and isotropy of the Fermi velocity and thus in the density of states. However, the anisotropic renormalization of the Fermi velocity due to strain remains a second order effect, neglected in this work \cite{choi2010effects}. At last, the electrostatic barrier effect due to distinct capacitances between the graphene regions on the top and bottom of the crenelation, which is otherwise included in our model, is too weak ($<$7 percent) to be responsible on its own for a $\Delta V_{strain}$ of 163 meV \cite{Drienovsky2018, Solís2016}. In the course of paper preparation, we became aware of a recent work addressing similar strained junction physics (arXiv:2312.00177).

\section{\label{sec:conclusion}conclusion}
In conclusion, using transport measurements at low temperature and microscopic modeling, we have investigated the Klein tunneling transport of ballistic electrons across a series of strain-induced barriers. We have shown the existence of a large and broad ancillary resistance peak at a positive energy that originates from nearly equivalent contribution of strain-induced scalar and pseudo-vector potentials in strained barriers. Our study also reveals the quantum interferences of ballistic electrons between strained graphene barriers. Our platform based on crenellated hBN-graphene heterostructures paves the way toward the realization of graphene quantum strain devices. It comprises the realization of high on/off transistors driven by a simple gate voltage resulting from the quantum nature of the transport without the need of a bandgap, and the implementation of valleytronics (filters and polarizers) as  electrons belonging to each valley in our strained device are collimated by strained barriers in different directions \cite{Yu2022}. Moreover, in our hBN-graphene crenellated heterostructure, the charge carriers undergo a sub-micron periodic angular motion, which should lead to the emission of electromagnetic radiation in the THz spectral range, paving the way towards the development of integrated synchrotron-like THz emitters based on 2D materials. For basic science, this platform gives new opportunities to develop strain engineering in a wide variety of 2D materials and Van der Waals heterostructures such as in twisted bilayer graphene or in transition metal dichalcogenides \cite{Peng2020} with the aim to manipulate their electronic and optical properties and explore new physical phenomena. 

\begin{acknowledgments}
The authors thank J. Tignon, J. Torres, F. Teppe, P. Legagneux, P. Seneor, B. Dlubak and M.B. Martin for fruitful discussions. \textbf{Funding} This project has received funding from the European Research Council (ERC) under the European Union's Horizon 2020 research and innovation program (Grant Agreement No. 820133). This research is supported by a public grant overseen by the French National Research Agency (ANR) as grants STEM2D (ANR-19-CE24-0015).
\end{acknowledgments}

\appendix

\section{Device fabrication}
To fabricate the crenellated hBN-graphene heterostructure, we transferred a 62 nm-thick exfoliated hBN flake on a silicon substrate with 500 nm oxide by using the
polymer-based dry pick-up and transfer technique. We etched trenches in the hBN flake of 35 nm deep by using e-beam lithography and reactive ion etching. We used also the polymer-based dry pick-up and transfer technique to assemble a thin hBN flake (8 nm) and exfoliated graphene and to transfer them onto the patterned hBN.  Afterwards, we contacted the crenellated graphene with 1D “edge contacts” to design source and drain electrodes by using e-beam lithography, reactive ion etching and the deposit of 5/50 nm of Cr/Au. To access the silicon that is used as a back gate, we patterned an electrical access through SiO$_2$ by using laser lithography, reactive ion etching and the deposit of 5/180 nm of Cr/Au. At the end, the crenellated graphene transistor is L=12 $\mu$m long and W=8 $\mu$m large and consists of 10 crenellations of length l=1 $\mu$m (i.e. with a 1 µm-periodicity) and of 2 unstrained graphene region of 1 $\mu$m length at the ends.

\section{Measurements}
AFM measurements were performed in air using silicon cantilevers operated in tapping mode. The Raman spectroscopy measurements were performed with a Renishaw inVia Raman microscope with a 100× objective lens and at an excitation wavelength of 532 nm. For transport measurements, the sample was cooled down in a variable-temperature (4-300 K) liquid 4He cryostat. An AC voltage of 1mV amplitude at a frequency f = 77 Hz and a DC voltage V$_{ds}$ were applied between the source and drain electrodes whereas a DC voltage V$_g$ was applied to the gate. The AC current signal was measured with a lock-in amplifier. 

\section{Diffusive contribution modeling}
To model the diffusive contribution to the resistance of our device, we used a capacitance calculated with the average distance to the gate, C$_{av}$=6.1x10$^{-9}$ F.cm$^{-2}$. The charge carrier density $n$ is then given by: $\vert n \vert = \sqrt{\left( \frac{C (V_g - V_0)}{e} \right)^2 +  n_0^2}$, where $V_0$ is a shift of the charge neutrality point due to intrinsic doping of the sample, $n_0$, a residual charge carrier density and $C$ the gate capacitance. The diffusive contribution in unstrained graphene regions was calculated from fits of the charge neutrality point peak according to $R=R_{c}+L/(W\lvert n\rvert e\mu)$, with R$_{c}$ the contact resistance, $\mu$ the the electron mobility over the entire structure, including unstained areas where mobility is assumed to be high, as evidenced by the observation of Fabry-Perot oscillations in the R$_{ds}$-V$_{g}$ characteristic, and strained barriers, and $n$ the charge density of the graphene. We found $R_{contact} =$ 4288 $\mathrm{\Omega}$, and  $n_0 = 4.5 \cdot 10^{11}~\mathrm{cm^{-2}}$ which is the order of magnitude expected for such graphene flake. The overall electronic mobility is 3604 $\mathrm{cm^{-2}.V^{-1}.s^{-1}}$ if we consider the total length of the device (\textit{i.e} 12 $\mathrm{\mu}$m). It should be pointed out that the electron mobility value includes, in addition to the presumed high-mobility unconstrained regions, 10 strained barriers of similar length that strongly disrupt carrier transport.
\\

\nocite{*}

\bibliography{Main_Send}

\end{document}